\documentclass[12pt,a4paper]{article}
\usepackage{epsfig}
\begin{document}
\title{Top quark chromomagnetic dipole moment in the littlest Higgs
model with T-parity }
\author{Li Ding and Chong-Xing Yue\\
{\small Department of Physics, Liaoning  Normal University, Dalian
116029, P. R. China}
\thanks{E-mail:cxyue@lnnu.edu.cn}}
\date{\today}

\maketitle
\begin{abstract}
The littlest Higgs model with T-parity, which is called $LHT$ model,
predicts the existence of the new particles, such as heavy top
quark, heavy gauge bosons, and mirror fermions. We calculate the
one-loop contributions of these new particles to the top quark
chromomagnetic dipole moment $(CMDM)$ $\Delta K$. We find that the
contribution of the $LHT$ model is one order of magnitude smaller
than the standard model prediction value.

\vspace{1cm}

\end{abstract}

\newpage
\noindent{\bf 1. Introduction}

The standard model $(SM)$ is an excellent low energy description of
the elementary particles. The next generation of high energy
colliders planned or under construction will test the $SM$ with high
precision and will explore higher energies in the search of new
physics. It is well known that new physics may manifest itself in
two ways: through direct signals involving the production of new
particles or by departures from the $SM$ predictions for the known
particles. In some cases, indirect effects can give evidence of new
physics beyond the $SM$ before new particles are discovered. Thus,
studying corrections of new physics to observables are important as
well.

The top quark, with a mass of the order of the electroweak scale
$m_t$ $\sim$ 172 $GeV$ [1], is the heaviest particle yet discovered
and might be the first place in which the new physics effects could
be appeared. The correction effects of new physics on observables
for top quark are often more important than for other fermions. Top
quarks will be copiously produced at the $CERN$ Large Hadron
Collider $(LHC)$ and later International Linear Collider $(ILC)$.
For example, 80 millions top quark pairs will be produced at the
$LHC$ per year [2]. This number will increase by one order of
magnitude with the high luminosity option. With such large samples,
precise measurements of its couplings will be available to test new
physics beyond the $SM$.

Anomalous top quark couplings can affect top production and decay at
high energy colliders as well as precisely measured quantities with
virtual top contributions. In particular, new contributions to the
top quark coupling $gt\bar{t}$, which is considered in this paper,
can produce significant contributions to the total and differential
cross sections of top-pair production at hadron colliders [3, 4]. If
the new physics, which can give rise to new contributions to the
coupling $gt\bar{t}$, is at the $TeV$ scale, then the anomalous
coupling $gt\bar{t}$ can be parameterized by the chromomagnetic
dipole moment $(CMDM)$ $\Delta K$ of the top quark, which is defined
as:
\begin{equation}
L=i(\frac{\Delta
K}{2})\frac{g_{s}}{2m_{t}}\bar{t}\sigma_{\mu\nu}q^{\nu}T^{a}tG^{\mu,a},
\end{equation}
where $\sigma_{\mu\nu}=\frac{i}{2}[\gamma_{\mu},\gamma_{\nu}]$,
$g_{s}$ and $T^{a}$ are the $SU(3)_{C}$
 coupling constant and generators, respectively. $q^{\nu}$ is the gluon momentum. Within the $SM$,
 the top quark $CMDM$ $\Delta K$
 vanish at the tree-level but can be generated at the one-loop level [5]. Recently, Ref.[6] has
 calculated the contributions of some new physics models to $\Delta K$ and they find that the
  topcolor assisted technicolor models can make
 its absolute value reach 0.01, which is of the order of the sensibility of the $LHC$.
  Motivited by the fact that the little Higgs theory [7]
 is one of the important candidates of the new physics beyond the
 $SM$ and $\Delta K$ would be more easily probed at the $LHC$ than at the Tevatron,
 in this paper we will reconsider the top quark $CMDM$
 $\Delta K$ in the context of the littlest Higgs model with T-party (called $LHT$ model)
 [8] and see whether its contributions to $\Delta K$ can be detected at the $LHC$.

 In the rest of this paper, we will give our results in detail. The
 essential features of the $LHT$ model, which are related our calculation, are reviewed in
 section 2. In section 3, the contributions of the new particles to the top quark $CMDM$ $\Delta K$
 are calculated. Conclusions are given in section 4.

\noindent{\bf 2. The essential features of the \textbf{$LHT$} model}

Little Higgs models are proposed as an alternative solution to the
hierarchy problem of the $SM$, which provide a possible kind of
electroweak symmetry breaking $(EWSB)$ mechanism accomplished by a
naturally light Higgs boson [7]. In this kind of models, the Higgs
boson is a pseudo-Goldstone boson and its mass is protected by a
global symmetry and quadratic divergence cancelations are due to
contributions from new particles with the same spin as the $SM$
particles. The $LHT$ model [8] is one of the attractive little Higgs
models. In order to satisfy the electroweak precision constraints by
avoiding tree-level contributions of the new particles and restoring
the custodial $SU(2)$ symmetry, a discrete symmetry, called
T-parity, is introduced in the $LHT$ model. Under T-parity, particle
fields are divided into T-even and T-odd sectors. The T-even sector
consists of the $SM$ particles and a heavy top $T_{+}$, while the
T-odd sector contains heavy gauge bosons $(B_{H}, Z_{H},
W_{H}^{\pm})$, a scalar triplet $(\Phi)$, and the so-called mirror
fermions.

The $LHT$ model is based on a $SU(5)/SO(5)$ global symmetry breaking
pattern, which gives rise to 14 Goldstone bosons:

\begin{equation}
\Pi=\left(
\begin{array}{ccccc}-\frac{\omega^{0}}{2}-\frac{\eta}{\sqrt{20}}
&-\frac{\omega^{+}}{\sqrt{2}}& -i\frac{\pi^{+}}{\sqrt{2}}&
-i\phi^{++} &
-i\frac{\phi^{+}}{\sqrt{2}} \\
-\frac{\omega^{-}}{\sqrt{2}}&\frac{\omega^{0}}{2}-\frac{\eta}{\sqrt{20}}
&\frac{\nu+h+i\pi^{0}}{2}&-i\frac{\phi^{+}}{\sqrt{2}}&\frac{-i\phi^{0}+\phi^{P}}{\sqrt{2}}
\\i\frac{\pi^{-}}{\sqrt{2}}&\frac{\nu+h-i\pi}{2}&\sqrt{4/5}\eta&
-i\frac{\pi^{+}}{\sqrt{2}}&\frac{\nu+h+i\pi}{2}\\
i\phi^{--}&i\frac{\phi^{-}}{\sqrt{2}}&i\frac{\pi^{-}}{\sqrt{2}}
&-\frac{\omega^{0}}{2}-\frac{\eta}{\sqrt{20}}&-\frac{\omega^{-}}
{\sqrt{2}}\\i\frac{\phi^{-}}{\sqrt{2}}&\frac{i\phi^{0}+\phi^{P}}{\sqrt{2}}
&\frac{\nu+h-i\pi}{2}&-\frac{\omega^{+}}{\sqrt{2}}&\frac{\omega^{0}}{2}-\frac{\eta}{\sqrt{20}}
\end{array}\right).
\end{equation}
Where it consists of a doublet $H$ and a triplet $\Phi$ under the
unbroken $SU(2)_{L}$$\times$$U(1)_{Y}$ group which are given by:
\begin{equation}
H=\left(
\begin{array}{c}-i\frac{\pi^{+}}{\sqrt{2}}\\\frac{\nu+h+i\pi^{0}}{2}\end{array}\right),
\hspace{1.5cm} \Phi=\left(
\begin{array}{cc}-i\phi^{++}&-i\frac{\phi^{+}}{\sqrt{2}}\\
-i\frac{\phi^{+}}{\sqrt{2}}&\frac{-i\phi^{0}+\phi^{P}}{\sqrt{2}}\end{array}\right).
\end{equation}
 Here $h$ is the physical Higgs field and $\nu = 246 GeV$ is the
 electroweak scale. The fields $\eta$ and $\omega$ are eaten by heavy
 gauge bosons when the $[SU(2) \times U(1)]^{2}$ gauge group is
 broken down to $SU(2)_{L} \times U(1)_{Y}$, whereas the $\pi$
 fields are absorbed by the $SM$ gauge bosons $W$ and $Z$ after $EWSB$.
 The fields $h$ and $\Phi$ remained in the particle spectrum are T-even and
 T-odd, respectively.

 After taking into account $EWSB$, at the order of $\nu^{2}/f^{2}$,
 the masses of the T-odd set of the $SU(2) \times U(1)$ gauge bosons
 are given as:
 \begin{equation}
M_{B_{H}}=\frac{g'f}{\sqrt{5}}[1-\frac{5\nu^{2}}{8f^{2}}],\hspace{1cm}
M_{Z_{H}}=M_{W_{H}}=gf[1-\frac{\nu^{2}}{8f^{2}}].
\end{equation}
Where $f$ is the scale parameter of the gauge symmetry breaking of
the $LHT$ model. $g'$ and $g$ are the $SM$ $U(1)_{Y}$ and
$SU(2)_{L}$ gauge coupling constants, respectively. Because of the
smallness of $g'$, the gauge boson $B_{H}$ is the lightest T-odd
particle, which can be seen as an attractive dark matter candidate
[9].

 In order to cancel the quadratic divergence of the Higgs mass
 coming from top loops, an additional heavy top quark $T_{+}$ need
 to be introduced, which is T-even and transforms as a singlet under
 $SU(2)_{L}$. Then the implementation of T-parity requires also a
 T-odd partner $T_{-}$, which is an exact singlet under $SU(2)_{1} \times
 SU(2)_{2}$. Their masses are:
 \begin{equation}
M_{T_{+}}=\frac{f}{\nu}\frac{m_{t}}{\sqrt{X_{L}(1-X_{L})}}[1+\frac{\nu^{2}}{f^{2}}
(\frac{1}{3}-X_{L}(1-X_{L}))],
\end{equation}
 \begin{equation}
M_{T_{-}}=\frac{f}{\nu}\frac{m_{t}}{\sqrt{X_{L}}}[1+\frac{\nu^{2}}{f^{2}}(\frac{1}{3}
-\frac{1}{2}X_{L}(1-X_{L}))].
\end{equation}
Where $X_{L}=\lambda_{1}^{2}/(\lambda_{1}^{2}+\lambda_{2}^{2})$ is
the mixing parameter between the $SM$ top quark $t$ and the new top
quark $T_{+}$, in which $\lambda_{1}$ and $\lambda_{2}$ are the
Yukawa coupling parameters.

   To avoid severe constraints and simultaneously implement T-parity,
it is need to double the $SM$ fermion doublet spectrum [8, 10, 11].
The T-even combination is associated with the $SU(2)_{L}$ doublet,
while the T-odd combination is its T-parity partner. The masses of
the T-odd fermions can be written in a unified manner as:

\begin{equation}
M_{F_{i}}=\sqrt{2}k_{i}f,
\end{equation}
where $k_{i}$ are the eigenvalues of the mass matrix $k$ and their
values are generally dependent on the fermion species $i$.

The mirror fermions (T-odd quarks and T-odd leptons) have new flavor
violating interactions with the $SM$ fermions mediated by the new
gauge bosons $(B_{H},W_{H}^{\pm}$, or $Z_{H})$, which are
parameterized by four $CKM$-$like$ unitary mixing matrices, two for
mirror quarks and two for mirror leptons [12, 13, 14]:
\begin{equation}
V_{Hu},\hspace*{0.2cm}V_{Hd},\hspace*{0.2cm}V_{Hl},\hspace*{0.2cm}V_{H\nu}.
\end{equation}
They satisfy:
\begin{equation}
V_{Hu}^{+}V_{Hd}=V_{CKM},\hspace*{0.6cm}V_{H\nu}^{+}V_{Hl}=V_{PMNS}.
\end{equation}
Where the $CKM$ matrix $V_{CKM} $ is defined through flavor mixing
in the down-type quark sector, while the $PMNS$ matrix $V_{PMNS} $
is defined through neutrino mixing.

From above discussions, we can see that, because of the presence of
new mixing matrices, the $LHT$ model might generate significant
effects on flavor observables and provide a picture of flavor
violating processes at scales above the electroweak scale that
differs dramatically from the $SM$ one. It has been shown that the
departures from the $SM$ expectations in the quark and lepton
sectors can be very large [11, 12, 13, 14, 15, 16]. In this paper,
we will focus our attention on the contributions of the $LHT$ model
to the top quark $CMDM$ $\Delta K$.

\noindent{\bf 3. The contributions of the \textbf{$LHT$} model to
the top quark $CMDM$ $\Delta K$}

 In the $LHT$ model, the top-quark $CMDM$ $\Delta K$ arises from loops containing the
 heavy top quarks $T_{+}$ and $T_{-}$, T-odd fermions $u_{H}^{i}$ and $d_{H}^{i}$, and the new
  gauge bosons $B_{H}, Z_{H}, W_{H}^{\pm}$
 with their respective would be Goldstone bosons. The relevant Feynman diagrams are shown in Fig.1.

\begin{figure}[htb]
\begin{center}
\epsfig{file=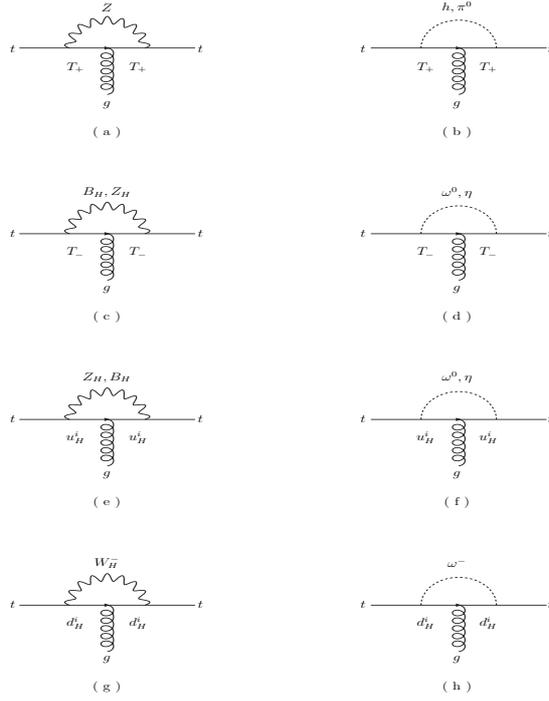,width=280pt,height=295pt}
 \vspace{-0.4cm}\caption{The Feynman diagrams contributing to $\Delta K$ in the $LHT$ model}
\end{center}
\end{figure}

Using the relevant Feynman rules given in Refs.[10, 12], we can
calculate the contributions of the new particles predicted by the
$LHT$ model to the top quark $CMDM$ $\Delta K$. Then the top quark
$CMDM$ $\Delta K$ are obtained as a sum of the contributions coming
from these new particles in the $LHT$ model.

 \begin{equation}
\Delta K=\Delta K_{T_{+}}+\Delta K_{T_{-}}+\Delta K_{u_{H}}+\Delta
K_{d_{H}}
\end{equation}

 \vspace{40cm}
with
 \begin{eqnarray}
\Delta K_{T_{+}}&=&-\frac{\alpha m_{t}^{2}\nu^{2}X_{L}^{2}}{4\pi
S_{W}^{2}C_{W}^{2}f^{2}}\int_{0}^{1}dx\int_{0}^{1-x}dy
\frac{(x+y-2)(x+y-1)}{m_{t}^{2}(x+y)^{2}+(x+y)(M_{T_{+}}^{2}
-M_{Z}^{2}-m_{t}^{2})+M_{Z}^{2}}-\frac{m_{t}^{3}}{8\pi^{2}\nu^{2}}
\nonumber\\&&\int_{0}^{1}dx\int_{0}^{1-x}dy\frac{(x+y)
\{\frac{2M_{T_{+}}\nu}{f}\sqrt{X_{L}(1-X_{L})}+m_{t}(x+y-1)[(1-X_{L})^{2}
\frac{\nu^{2}}{f^{2}}+\frac{X_{L}}{1-X_{L}}]}{m_{t}^{2}(x+y)^{2}
+(x+y)(M_{T_{+}}^{2}-m_{h}^{2}-m_{t}^{2})+m_{h}^{2}}
 \nonumber\\&&-\frac{\alpha
m_{t}^{3}X_{L}^{2}}{8\pi
m_{Z}^{2}S_{W}^{2}C_{W}^{2}}\int_{0}^{1}dx\int_{0}^{1-x}dy
\frac{(x+y)[m_{t}(x+y-1)(\frac{1}{X_{L}(1-X_{L})}+\frac{\nu^{2}}{f^{2}})+\frac{2M_
{T_{+}}\nu}{f\sqrt{X_{L}(1-X_{L})}}]}{m_{t}^{2}(x+y)^{2}+(x+y)(M_{T_{+}}^{2}
-m_{\pi}^{2}-m_{t}^{2})+m_{\pi}^{2}},\nonumber\\
\hspace*{0.3cm}
\end{eqnarray}

 \begin{eqnarray}
\Delta K_{T_{-}}&=&-\frac{4\alpha m_{t}X_{L}}{25\pi
C_{W}^{2}}\int_{0}^{1}dx\int_{0}^{1-x}dy\frac{(x+y-1)
[m_{t}(x+y-2)(X_{L}\frac{\nu^{2}}{f^{2}}+1)-2M_{T_{-}}
\sqrt{X_{L}}\frac{\nu}{f}]}{m_{t}^{2}(x+y)^{2}+(x+y)(M_{T_{-}}^{2}
-M_{B_{H}}^{2}-m_{t}^{2})+M_{B_{H}}^{2}}\nonumber\\
&&-\frac{m_{t}^{4}}{2\pi^{2}}\int_{0}^{1}dx\int_{0}^{1-x}dy[\frac{\nu^{2}}{16f^{4}}\frac{(x+y)(x+y-1)}
{m_{t}^{2}(x+y)^{2}+(x+y)(M_{T_{-}}^{2}-M_{\omega^{0}}^{2}-m_{t}^{2})+M_{\omega^{0}}^{2}}\nonumber\\
&&+\frac{1}{5\nu^{2}}\frac{(x+y)(x+y-1)}
{m_{t}^{2}(x+y)^{2}+(x+y)(M_{T_{-}}^{2}-M_{\eta}^{2}-m_{t}^{2})+M_{\eta}^{2}}],
      \hspace*{4.8cm}
\end{eqnarray}

\begin{eqnarray}
\Delta K_{u_{H}}&=&-\sum_{i=1,2,3}\frac{\alpha
m_{t}^{2}[(V_{Hu})_{i3}]^{2}}{4\pi
S_{W}^{2}C_{W}^{2}}\int_{0}^{1}dx\int_{0}^{1-x}dy[\frac{C_{W}^{2}(x+y-2)(x+y-1)}{m_{t}^{2}(x+y)^{2}+(x+y)
(M_{u_{H}^{i}}^{2}-M_{Z_{H}}^{2}-m_{t}^{2})+M_{Z_{H}}^{2}}\nonumber\\
&&+\frac{S_{W}^{2}}{25}\frac{(x+y-2)(x+y-1)}{m_{t}^{2}(x+y)^{2}+(x+y)(M_{u_{H}^{i}}^{2}-M_{B_{H}}^{2}
-m_{t}^{2})+M_{B_{H}}^{2}}]
-\sum_{i=1,2,3}\frac{m_{t}^{2}[(V_{Hu})_{i3}]^{2}}{32\pi^{2}f^{2}}\nonumber\\
&&\int_{0}^{1}dx\int_{0}^{1-x}dy
[\frac{(x+y)[2M_{u_{H}^{i}}^{2}+(x+y-1)(M_{u_{H}^{i}}^{2}+m_{t}^{2})]}{m_{t}^{2}(x+y)^{2}+(x+y)
(M_{u_{H}^{i}}^{2}-M_{\omega^{0}}^{2}-m_{t}^{2})+M_{\omega^{0}}^{2}}\nonumber\\
&&+\frac{1}{5}\frac{(x+y)[2M_{u_{H}^{i}}^{2}+(x+y-1)
(M_{u_{H}^{i}}^{2}+m_{t}^{2})]}{m_{t}^{2}
(x+y)^{2}+(x+y)(M_{u_{H}^{i}}^{2}-M_{\eta}^{2}-m_{t}^{2})+M_{\eta}^{2}}],
\hspace*{0.8cm}
\end{eqnarray}

\begin{eqnarray}
\Delta K_{d_{H}}&=&-\sum_{i=1,2,3}\frac{\alpha
m_{t}^{2}[(V_{Hu})_{i3}]^{2}}{2\pi
S_{W}^{2}}\int_{0}^{1}dx\int_{0}^{1-x}dy\frac{(x+y-2)(x+y-1)}{m_{t}^{2}(x+y)^{2}+(x+y)
(M_{d_{H}^{i}}^{2}-M_{W_{H}}^{2}-m_{t}^{2})+M_{W_{H}}^{2}}\nonumber\\
&&-\sum_{i=1,2,3}\frac{m_{t}^{2}[(V_{Hu})_{i3}]^{2}}{16\pi^{2}f^{2}}\int_{0}^{1}dx\int_{0}^{1-x}dy
[\frac{(x+y)[(x+y-1)(M_{d_{H}^{i}}^{2}+m_{t}^{2})+2M_{d_{H}^{i}}^{2}]}{m_{t}^{2}(x+y)^{2}
+(x+y)(M_{d_{H}^{i}}^{2}-M_{\omega^{-}}^{2}-m_{t}^{2})+M_{\omega^{-}}^{2}}.\nonumber
\\\hspace{0.2cm}
\end{eqnarray}
Where $S_{W}=\sin \theta_{W}$, $\theta_{W}$ is the Weinberg angle.
$m_{\pi}$, $ M_{\omega},$ and $M_{\eta}$ are the masses of the would
be Goldstone bosons $\pi, \omega,$ and $\eta$, respectively. Similar
with Ref.[11], we will assume $m_{\pi^{0}}=M_{Z}$,
 $M_{\omega^{0}}=M_{Z_{H}}$, $M_{\omega^{\pm}}=M_{W_{H}^{\pm}}$ and
$M_{\eta}=M_{B_{H}}$ in our numerical calculation. In above
equations, we only consider the Feynman rules at the order of
$\nu/f$, which give contributions to the top quark $CMDM$ $\Delta K$
at the $\nu^{2}/f^{2}$ level. At order $\nu^{2}/f^{2}$ calculation,
we have neglected the contributions of the heavy gauge boson $Z_{H}$
in above equations.

\begin{figure}[htb]
\begin{center}
\epsfig{file=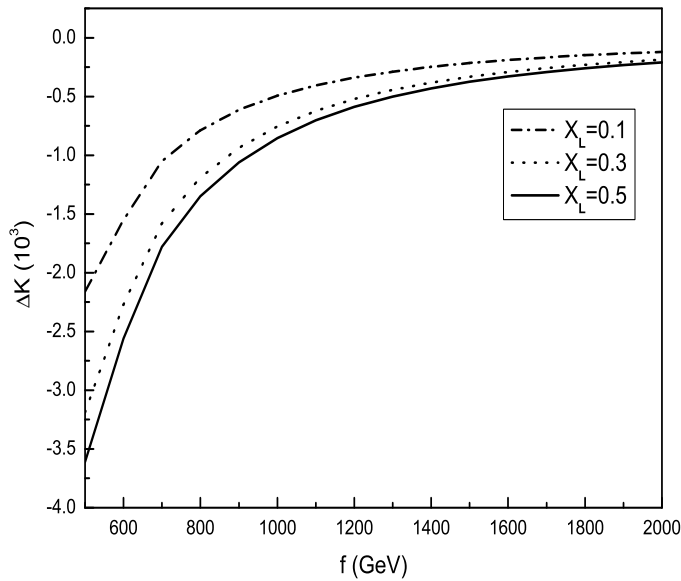,width=285pt,height=255pt}
 \vspace{-0.4cm}\caption{In case I, the top quark $CMDM$ $\Delta K$ as a function of the
 scale parameter $f$ \hspace*{2.0cm}for three values of the mixing parameter $X_{L}$.}
\end{center}
\end{figure}

From above equations, we can see that the free parameters which are
related to our analysis are $f, X_{L}, M_{u_{H}^{i}}, M_{d_{H}^{i}}$
and $V_{Hu}$. The matrix elements $(V_{Hu})_{ij}$ can be determined
through $V_{Hu}=V_{Hd}V_{CKM}^{+}$. The matrix $V_{Hd}$ can be
parameterized in terms of three mixing angles and three phases,
which can be probed by $FCNC$ processes in $K$ and $B$ meson
systems, as discussed in detail in Refs.[12, 14]. It is convenient
to consider several representative scenarios for the structure of
the $V_{Hd}$. To simply our calculation, we concentrate our study on
the following two scenarios for the structure of the mixing matrix
$V_{Hd}$:\\
\hspace*{0.6cm}case I:  \hspace{1.6cm}   $ V_{Hd}=I,\hspace{0.5cm} V_{Hu}=V_{CKM}^{+};$\\
\hspace*{0.6cm}case II:   \hspace{1.6cm}
$S_{23}^{d}=1/\sqrt{2},\hspace{0.3cm} S_{12}^{d}=0,\hspace{0.3cm}
S_{13}^{d}=0, \hspace{0.3cm}\delta_{12}^{d}=0,\hspace{0.3cm}
\delta_{23}^{d}=0,\hspace{0.3cm} \delta_{13}^{d}=0$.

It has been shown that, in both cases, the constraints on the mass
spectrum of the T-odd fermions are very relaxed [12, 14]. From
Eqs.(11)-(14), we can see that the contributions of the $LHT$ model
to the top quark $CMDM$ $\Delta K$ are mainly dependent on the scale
parameter $f$. Thus, we will assume
$M_{u_{H}^{i}}=M_{d_{H}^{i}}=400GeV (i=1, 2),
M_{u_{H}^{3}}=M_{d_{H}^{3}}=600GeV$ and take the scale parameter $f$
and the mixing parameter $X_{L}$ as free parameters in our numerical
estimation.

\begin{figure}[htb]
\begin{center}
\epsfig{file=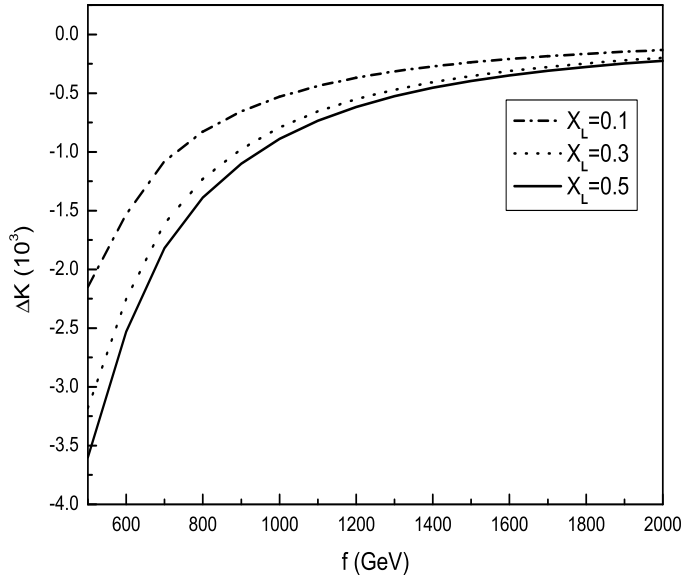,width=285pt,height=255pt}
 \vspace{-0.4cm}\caption{Same as Fig.2 but for case II.}
\end{center}
\end{figure}

 Our numerical results our summarized in Fig.2 and Fig.3. One can see from these figures that the top quark
 $CMDM$ $\Delta K$ is strongly dependent on the parameters $f$. The contribution
 of the $LHT$ model to $\Delta K$ for case I has not large deviations from that for
 case II. This means that the total contributions of T-odd fermions are not sensitive to
 the mixing matrix elements $(V_{Hu})_{ij}$. If we assume $f=500GeV-2000GeV$ and $0.1 \leq X_{L} \leq
 0.5$, then the values of $\Delta K$ are in the ranges of
$ -3.6\times10^{-3} \sim -1.23\times10^{-4}$ and
 $-3.61\times10^{-3}\sim
 -1.34\times10^{-4}$ for case I and case II, respectively.

 It is well known that the severe constraint on the top quark $CMDM$ $\Delta
 K$ comes from the process $b \rightarrow s \gamma$, which is as strong as
 that expected at the $LHC$. Ref.[5] has shown that the constraint
 given by the $CLEO$ data on $b \rightarrow s \gamma$ is about $-0.03 \leq \Delta K \leq
 0.01$. Our numerical results show that the contribution of the $LHT$ model to $\Delta
 K$ is consistent with this constraint. However, the contribution of the $LHT$ model to $\Delta K$
 is about one order of magnitude smaller than the $SM$ prediction value, which is difficult to be
 detected in near future $LHC$.

 \noindent{\bf 4. Conclusions }

 The $LHT$ model is one of the attractive little Higgs models, which
 provides a possible dark matter candidate. To simultaneously
 implement T-parity, the $LHT$ model introduces new mirror fermions.
 The flavor mixing in the mirror fermion sector give rise to a new
 source of flavor violation, which can generate contributions to
 some flavor observables.

 The $LHT$ model predicts the existence of new particles, such as
 new  heavy top quark ($T_{+}, T_{-}$), new gauge bosons ($B_{H}, W_{H}^{\pm},$ and
 $Z_{H}$) and T-odd fermions. In this paper, we have calculated the
 one-loop contributions of these new particles to the top quark
 $CMDM$ $\Delta K$. We find that the absolute value of $\Delta K$ is
 at order of $1 \times 10^{-3}$ in wide range of the parameter space,
 which is smaller than that given in the context of the $SM$ by one
 order of magnitude.

 The $LHC$ will become operational in next year. One of the primary
 goals for the $LHC$ is to determine the top quark properties and
 see whether any hint of non-$SM$ effects may be visible. The
 next-to-leading order cross section for top quark pair production
 at the $LHC$ with $\sqrt{s}=14TeV$ is calculated to be 833pb of
 which some references are given in Refs.[2, 17]. This will make it
 possible to precisely determine the top quark couplings and also
 offers an excellent chance to search for new physics at the $LHC$.
 Thus one might distinguish different new physics models via
precision  measurement of the top quark $CMDM$ at the $LHC$.

\vspace{1.0cm}

\noindent{\bf Acknowledgments}

This work was supported in part by the National Natural Science
Foundation of China under Grants No.10675057 and Foundation of
Liaoning  Educational Committee(2007T086).

\end{document}